\def\Roman#1{\uppercase\expandafter{\romannumeral#1}}
\documentclass[a4paper,14pt]{article}
\usepackage{amssymb,amsfonts}
\usepackage[mathscr]{eucal}

\title{Path integral representation of the quantum evolution in dynamical systems with a symmetry for the non-zero momentum level reduction}

\author{S.N.Storchak}

\begin{document}

\maketitle

\begin{abstract}
For the case of reduction onto the non-zero momentum level, in the problem of the path integral quantization of a scalar particle motion on a smooth compact Riemannian manifold with the given free isometric action of the compact semisimle Lie group, the path integral representation of the matrix Green's function, which describes the quantum evolution of the reduced motion, has been obtained.  
The integral relation between the path integrals representing the fundamental solutions of the parabolic differential equation defined on the total space of the principal fiber bundle and the linear parabolic system  of the differential  equations on the space of the sections of the associated covector bundle has been derived. 
\end{abstract}

\section{Introduction}
 There is a number of a remarkable properties in dynamical systems with a symmetry. 
The main property of these systems manifests itself in 
a relationship between an original system and another system  (a reduced one) 
obtained from the original system after
``removing''  the   group degrees of freedom. 

One of the system of this class is the dynamical system which 
describes a motion  of a scalar particle on a smooth compact finite-dimensional Riemannian manifold with a  given free isometric smooth action of a  semisimple compact Lie group. 
 In fact, the original motion of the particle takes place    on the total space of a principal fiber bundle,  and the reduced motion --- on the orbit space of this bundle. 
 
This system bears close resemblance to the gauge field models,
where the reduced evolution is given on the orbit space of a gauge group action. That is why 
a great deal of attention has been devoted to the quantization of the finite-dimensional system related to the particle motion on a manifold with a group action \cite{Tanim,Kunsht,Falk,Lott}. 

 In gauge theories,  the motion on the orbit space is described  in terms of the gauge fields that are restricted to a gauge surface. Moreover,  a description of this motion  is only possible by means of  dependent variables.
 
Such a description is used in a heuristic method of the path integral quantization of the gauge fields proposed by Faddeev and Popov \cite{Faddeev}. 
However, at present, it is not even quite clear   how to define correctly the path integral measure on the space of the gauge fields. 
Therefore, in order to establish 
the final validity of the method 
it would be desirable to  carry out its additional investigation from the standpoint of a general approach developed in  the integration theory. 
There is a hope that it gives us an answer on yet unsolved questions of the Faddeev--Popov method.

As a first step in that direction, 
it was  studied  the path integral reduction 
in the aforementioned finite-dimensional  dynamical system \cite{Storchak_1}.
We have used the methods of the stochastic process theory
for definition of a path integral measure and in order to  study   the path integral transformation under the reduction. That is, we  dealt with  diffusion on a manifold with a given group action and with the path integral representation of the solution of the backward Kolmogorov equation.

Path integral reduction is based on the 
 separation of the variables or, in other words,   on the factorization of the original path integral measure into the `group' measure 
and the measure that is given  on the orbit space. In our papers, it was fulfilled   with the help of the nonlinear filtering stochasic differential equation. 
Note that a similar approach to the measure factorization was developed in \cite{Elworthy_2}.
Also,  the questions related to the factorization  have been studied in \cite{Paycha}.

As a result of the reduction,
  the integral relation between  the  wave  functions of the corresponding   `quantum' evolutions (the reduced and original diffusions) 
was obtained.

It was found that the Hamilton operator of the reduced dynamical system (the differential generator of the stochastic process) has an extra potential term. This  term comes from  from the reduction Jacobian.

In \cite{Storchak_2},  the path integral reduction 
has been considered
in the case 
when the reduced motion is described   in terms of the dependent variables.
 
  As in gauge theories, we have  suggested  that  the principal bundle 
is a trivial one. Then, in the principal fiber bundle, there is a global cross-section. The cross-section   may be determined  with the  choice of the special gauge surface. The  evolution on this gauge surface serves for description of the corresponding  reduced evolution on the orbit space.  

In this paper we will study the case of the  non-zero momentum level reduction in the path integral for the discussed  finite-dimensional dynamical system. 
The path integral, which describes the evolution of the reduced motion on the orbit space, will be represent the fundamental solution of the linear parabolic system of the differential equations.

\section{Definitions}
In our papers \cite{Storchak_1}, we have considered the diffusion of a scalar particle on a smooth compact Riemannian manifold $\cal P$. The backward Kolmogorov equation
for the  original  diffusion 
was as follows
 \begin{equation}
\left\{
\begin{array}{l}
\displaystyle
\left(
\frac \partial {\partial t_a}
+\frac 12\mu ^2\kappa \triangle
_{\cal P}(p_a)+\frac
1{\mu ^2\kappa m}
V(p_a)\right){\psi}_{t_b} (p_a,t_a)=0\\
{\psi}_{t_b} (p_b,t_b)=\phi _0(p_b),
\qquad\qquad\qquad\qquad\qquad (t_{b}>t_{a}),
\end{array}\right.
\label{1}
\end{equation}
where $\mu ^2=\frac \hbar m$ ,
$\kappa $ is a real positive parameter,
$\triangle _{\cal P}(p_a)$  is a Laplace--Beltrami operator on $\cal P$, and $V(p)$ is a group--invariant potential term. In a chart with the coordinate functions $Q^A=
{\varphi}^A (p)$, $p\in {\cal P}$, the Laplace -- Beltrami operator is written as
\[
\triangle _{\cal P}(Q)=G^{-1/2}(Q)\frac \partial
{\partial Q^A}G^{AB}(Q)
G^{1/2}(Q)\frac\partial {\partial Q^B},
\]
with $G=det (G_{AB})$,  $G_{AB}(Q)=G(\frac{\partial}{\partial Q^A},\frac{\partial}{\partial Q^B})$.

In accordance with  the theory developed by Daletskii and Belopolskaya  \cite{Daletskii}, the solution of (\ref{1}) is given by the global semigroup which is a limit (under the refinement of the subdivision of the time interval)
 of a superposition of the local semigroups 
\begin{equation}
\psi _{t_b}(p_a,t_a)=U(t_b,t_a)\phi _0(p_a)=
{\lim}_q {\tilde U}_{\eta}(t_a,t_1)\cdot\ldots\cdot
{\tilde U}_{\eta}(t_{n-1},t_b)
\phi _0(p_a).
\label{5}
\end{equation}
Each  local semigroup is determined by the path integrals with
 the
 integration measures defined by the local representatives
 $\eta ^A (t)$ of the global stochastic process $\eta (t)$.
The local stochastic process $\eta ^{A}(t)$ are  given by  the solutions of  the following stochastic differential equations:
\begin{equation}
d\eta ^A(t)=\frac12\mu ^2\kappa G^{-1/2}\frac \partial {\partial
Q^B}(G^{1/2}G^{AB})dt+\mu \sqrt{\kappa }{\mathfrak
X}_{\bar{M}}^A(\eta
(t))
dw^{
\bar{M}}(t),
\label{3}
\end{equation}
where the matrix ${\mathfrak X}_{\bar{M}}^A$ is defined 
by the local equality  
$\sum^{n_{P}}_{\bar{
{\scriptscriptstyle K}}
\scriptscriptstyle =1}
{\mathfrak X}_{\bar{K}}^A{\mathfrak X}_
{\bar{K}}^B=G^{AB}$.\\
 (We denote the Euclidean indices by
 over--barred indices.) 

Therefore, the behavior of the global semigroup  (\ref{5}) is completely defined by these stochastic differential equations.
 The global semigroup 
can be written symbolically as follows 
\begin{eqnarray}
{\psi}_{t_b} (p_a,t_a)&=&{\rm E}\Bigl[\phi _0(\eta (t_b))
\exp \{\frac 1{\mu
^2\kappa m}\int_{t_a}^{t_b}V(\eta (u))du\}\Bigr]\nonumber\\
&=&\int_{\Omega _{-}}d\mu ^\eta (\omega )
\phi _0(\eta (t_b))\exp \{\ldots \},
\label{2}
\end{eqnarray}
where  ${\eta}(t)$ is a global stochastic process on a manifold
$\cal
P$. $\Omega _{-}=\{\omega (t):\omega (t_a)=0,
\eta (t)=p_a+\omega (t)\}$  is the path space
 on this manifold. The path integral  
 measure ${\mu}^{\eta}$ is  defined by the probability
 distribution
of a stochastic process ${\eta}(t)$.
  
\subsection{Geometry of the problem}
Since in our case, there is a  free  isometric smooth action of a semisimple compact Lie group $\cal G$ on the original manifold $\cal P$, this manifold can be viewed 
as a total space of the principal fiber bundle $\pi : \cal P \to {\cal P}/{\cal G}=\cal M$. 

At the first step of the reduction procedure, we have transformed the original coordinates $Q^A$  given on a local chart of the manifold $\cal P$ for new coordinates 
$(Q^{\ast}{}^A,a^{\alpha})$ ($A=1,\ldots , N_{\cal P},N_{\cal P}=\dim {\cal P} ;{\alpha}=1,\ldots , N_{\cal G},N_{\cal G}=\dim {\cal G}$)
  related to the fiber  bundle. In order to meet 
a requirement of a  one-to-one mapping between  $Q^A$ and  $(Q^{\ast}{}^A,a^{\alpha})$, we are forced to introduce the additional constraints, ${\chi}^{\alpha}(Q^{\ast})=0$.

These constraints define the local submanifolds in the  manifold $\cal P$.
On the assumption that these local submanifolds (local sections) can be `glued' into the global manifold $\Sigma $,
  we come to a trivial principal fiber bundle $P({\cal M},\cal G)$.

We note that this bundle is locally isomorphic to the trivial bundle 
$\Sigma\times {\cal G}\to{\Sigma} $. It allows us to use the coordinates $Q^{\ast}{}^A$ for description of the evolution on the manifold $\cal M$.

If we replace the coordinate basis $(\frac{\partial}{\partial Q^A})$ for 
a new coordinate basis $(\frac{\partial}{\partial Q^{\ast}{}^A},\frac{\partial}{\partial a^{\alpha}})$, we get the following representation for  the original metric ${\tilde G}_{\cal A\cal B}(Q^{\ast},a)$ of the manifold $\cal P$: 
\begin{equation}
\left(
\begin{array}{cc}
G_{CD}(Q^{\ast})(P_{\perp})^{C}_{A}
(P_{\perp})^{D}_{B} & G_{CD}(Q^{\ast})(P_{\perp})^
{D}_{A}K^{C}_{\mu}\bar{u}^{\mu}_{\alpha}(a) \\
G_{CD}(Q^{\ast})(P_{\perp})^
{C}_{A}K^{D}_{\nu}\bar{u}^{\nu}_{\beta}(a) & {\gamma }_{\mu \nu }
(Q^{\ast})\bar{u}_\alpha ^\mu (a)\bar{u}_\beta ^\nu (a)
\end{array}
\right).
\label{8}
\end{equation}
To obtain this expression we have used   the right action of the group $\cal G $  on a manifold $\cal P$. 
It was given by functions $F^A(Q,a)$, performing an action, and their derivatives: $F^{C}_{B}(Q,a)\equiv \frac{\partial F^{C}}{\partial Q^{B}}(Q,a)$. 
For example, 
$G_{CD}(Q^{\ast})\equiv G_{CD}(F(Q^{\ast},e))$ is 
defined as
\[
G_{CD}(Q^{\ast})=F^{M}_C(Q^{\ast},a)F^N_D(Q^{\ast},a)
G_{MN}(F(Q^{\ast},a)),
\]
 ($e$ is an identity element  of the group $\cal G$). 
 In (\ref{8}), the Killing vector fields $K_{\mu}$ for the Riemannian metric 
$G_{AB}(Q)$
are also  taken on the submanifold $\Sigma \equiv\{{\chi}^{\alpha}=0\}$, i.e.  the components $K^A_{\mu}$ depend on $Q^{\ast}$.
By ${\gamma}_{\mu \nu}$, defined as ${\gamma}_{\mu \nu}
=K^{A}_{\mu}G_{AB}K^{B}_{\nu}$,
we denote 
the metric given on the orbit of the group action. 

The operator $P_{\perp}(Q^{\ast})$, which projects the vectors onto the tangent space to the gauge surface $\Sigma$, has the following form:
\[
(P_{\perp})^{A}_{B}=\delta ^{A}_{B}-{\chi}^{\alpha}_{B}
(\chi \chi ^{\top})^{-1}{}^{\beta}_{\alpha}(\chi ^
{\top})^{A}_{\beta},
\]
 $(\chi ^{\top})^{A}_{\beta}$ is a transposed matrix to the matrix $\chi ^{\nu}_{B}\equiv \frac{\partial \chi ^{\nu}}{\partial Q^B}$, 
$(\chi ^{\top})^{A}_{\mu}=G^{AB}{\gamma}_
{\mu \nu}\chi ^{\nu}_{B}.$

The pseudoinverse matrix ${\tilde G}^{\cal A\cal B}(Q^{\ast},a)$ to the matrix (\ref{8}) is determined by the equality
\begin{eqnarray*}
\displaystyle
{\tilde G}^{\cal A\cal B}{\tilde G}_{\cal B\cal C}=\left(
\begin{array}{cc}
(P_{\perp})^A_C & 0\\
0 & {\delta}^{\alpha}_{\beta}
\end{array}
\right).
\end{eqnarray*}  
It follows that ${\tilde G}^{\cal A\cal B}$ is equal to 
\begin{equation}
\displaystyle
\left(
\begin{array}{cc}
G^{EF}N^{C}_{E}
N^{D}_{F} & G^{SD}N^C_S{\chi}^{\mu}_D
(\Phi ^{-1})^{\nu}_{\mu}{\bar v}^{\sigma}_{\nu} \\
G^{CB}{\chi}^{\gamma}_C (\Phi ^{-1})^{\beta}_{\gamma}N^D_B
{\bar v}^{\alpha}_{\beta} & G^{CB}
{\chi}^{\gamma}_C (\Phi ^{-1})^{\beta}_{\gamma}
{\chi}^{\mu}_B (\Phi ^{-1})^{\nu}_{\mu}
{\bar v}^{\alpha}_{\beta}{\bar v}^{\sigma}_{\nu}
\end{array}
\right).
\label{9}
\end{equation}
The matrix  $(\Phi ^{-1}){}^{\beta}_{\mu}$  is inverse
to the Faddeev -- Popov matrix $\Phi $, which is given by
\[
(\Phi ){}^{\beta}_{\mu}(Q)=K^{A}_{\mu}(Q)
\frac{\partial {\chi}^{\beta}(Q)}{\partial Q^{A}}.
\]
In (\ref{9}), 
\[
N^{A}_{C}\equiv{\delta}^{A}_{C}-K^{A}_{\alpha }
(\Phi ^{-1}){}^{\alpha}_{\mu}{\chi}^{\mu}_{C}
\]
is a projection operator  with the following properties:
\[
N^{A}_{B}N^{B}_{C}=N^{A}_{C},\,\,\,\,\,N^A_BK^B_{\mu}=0,\,\,\,\,\,
(P_{\perp})^{\tilde A}_{B}N^{C}_{\tilde A}=
(P_{\perp})^{C}_{B},\,\,\,\,\,\,\,N^{\tilde A}_
{B}(P_{\perp})^{C}_{\tilde A}=N^{C}_{B}.
\]
The matrix  ${\bar v}^{\alpha}_{\beta}(a)$ is
an inverse matrix to
 matrix ${\bar u}^{\alpha}_{\beta}(a)$.
The $\det {\bar
u}^{\alpha}_{\beta}(a)$ is a density of a right invariant
measure
given on the group $\cal G$.

The determinant of the matrix (\ref{8}) is equal to
\begin{eqnarray*}
&&(\det {\tilde G}_{\cal A\cal B})=
\det G_{AB}(Q^{\ast})\det {\gamma}_{\alpha \beta}(Q^{\ast})
(\det {\chi}{\chi }^{\top})^{-1}(Q^{\ast})
(\det {\bar u}^{\mu}_{\nu}(a))^2\nonumber\\
&&\,\,\,\;\;\;\;\;\;\;\;\;\;\;\;\;\;\;\times(\det 
{\Phi}^{\alpha}_{\beta} (Q^{\ast}))^2
\det (P_{\perp})^C_B(Q^{\ast})\nonumber\\
&&\,\,\,\;\;\;\;\;\;\;\;\;\;\;\;\;\;\;=\det\Bigl((P_{\perp})^D_A \;G^{\rm H}_{DC}\,(P_{\perp})^C_B\Bigr)        \det {\gamma}_{\alpha \beta}\,\,(\det {\bar u}^{\mu}_{\nu})^2,  
\end{eqnarray*}
where
the ``horizontal metric'' $G^{\rm H}$
is defined by the relation $G^{\rm H}_{DC}={\Pi}^{\tilde D}_D\,{\Pi}^{\tilde C}_C\,G_{{\tilde D}{\tilde C}}$, in which   ${\Pi}^{ A}_B={\delta}^A_B-K^A_{\mu}{\gamma}^{\mu \nu}K^D_{\nu}G_{DB}$ is  the projection operator. (From the definition of ${\Pi}^{ A}_B$ it follows  that ${\Pi}^{ A}_L N^L_C={\Pi}^{ A}_C$ and ${\Pi}^L_BN^A_L=N^A_B$.)

Note also that $\det {\tilde G}_{\cal A\cal B}$ does not vanish only on the surface $\Sigma$.
On this surface $\det (P_{\perp})^C_B$ is equal to  unity.

\subsection{The semigroup on $\Sigma$ and its path integral representation} 
Transition to the bundle coordinates on $\cal P$ leads to the replacement of the local stochastic process ${\eta}^A_t$ for the process ${\zeta}^A_t=({Q_t^{\ast}}^A,a^{\alpha}_t)$.\footnote{This  phase space transformation of the stochastic processes does not change the path integral measures in the  evolution semigroups.}   
  Instead of the stochastic differential equation for the process 
${\eta}^A_t$  we get the system of equations for the processes ${Q_t^{\ast}}^A$ and $a^{\alpha}_t$:
 \begin{equation}
dQ_t^{*}{}^{\small A}={\mu}^2\kappa
\biggl(-\frac12
G^{EM}N^C_EN^B_M\,{}^{\rm H}{\Gamma}^A_{CB}+
j^{\small A}_{\Roman 1}+j^{\small A}_{\Roman
2}\biggr)dt +\mu\sqrt{\kappa}
N^A_C\tilde{\mathfrak X}^C_{\bar M}dw^{\bar M}_t,
\label{sde_q}
\end{equation}
\begin{eqnarray}
&&da_t^{\alpha}=-\frac12{\mu}^2\kappa\biggl[G^{RS}
  \tilde{\Gamma}^B_{RS}(Q^*){\Lambda}^{\beta}_B
  {\bar v}^{\alpha}_{\beta}
  +G^{RP}{\Lambda}^{\sigma}_R
  {\Lambda}^{\beta}_BK^B_{\sigma P}
  {\bar v}^{\alpha}_{\beta}
-G^{CA}N^M_C{\Lambda}^{\beta}_{AM}
  {\bar v}^{\alpha}_{\beta}
\nonumber\\
&&-G^{MB}{\Lambda}^{\epsilon}_M
  {\Lambda}^{\beta}_B{\bar v}^{\nu}_{\epsilon}
  \frac{\partial}
  {\partial a^{\nu}}
  \bigl({\bar v}^{\alpha}_{\beta}\bigr) 
  \biggr]dt
+\mu\sqrt{\kappa}
  {\bar v}^{\alpha}_{\beta}{\Lambda}^{\beta}_B
  \tilde{\mathfrak X}^B_{\bar M}dw_t^{\bar M}.
  \label{sde_a}
  \end{eqnarray}
In these equations,  ${\bar v}\equiv {\bar v}(a)$, and the
other coefficients depend on $Q^*$.

In equation (\ref{sde_q}), ${}^{\rm H}{\Gamma}^B_{CD}$ are the Christoffel symbols defined by the equality
\begin{equation}
G^{\rm H}_{AB}\,
{}^{\rm H}{\Gamma}^B_{CD}
=\frac12\left(G^{\rm H}_{AC,D}+
G^{\rm H}_{AD,C}-G^{\rm H}_{CD,A}
\right),
\label{sd6}
\end{equation}
in which by the derivatives we mean the following: 
$G^{\rm H}_{AC,D}\equiv 
\left.{{\partial G^{\rm H}_{AC}(Q)}\over
{\partial Q^D}}\right|_{Q=Q^{*}}$.
Also, by $j_{\Roman 1}$  we have denoted the mean curvature vector of the orbit space, and by $j_{\Roman 2}^A(Q^{*})$ --- the projection 
of the mean curvature vector of the orbit onto the submanifold $\Sigma $. This vector can be defined as
\begin{eqnarray}
  j_{\Roman
  2}^A(Q^{*})&=&-\frac12\,G^{EU}N^A_EN^D_U
  \left[{\gamma}^{\alpha\beta}G_{CD}
  ({\tilde{\nabla}}_{K_{\alpha}}
  K_{\beta})^C\right](Q^{\ast})\nonumber\\
&=&-\frac12\,N^A_C
  \left[{\gamma}^{\alpha\beta}
  ({\tilde{\nabla}}_{K_{\alpha}}
  K_{\beta})^C\right](Q^{\ast}),
\label{j2} 
\end{eqnarray}
where
\[
({\tilde{\nabla}}_{K_{\alpha}}
  K_{\beta})^C(Q^{\ast})= 
  K^A_{\alpha}(Q^{\ast})\left.\frac{\partial}
  {\partial Q^A}K^C_{\beta}(Q)\right |_{Q=Q^{\ast}}+
  K^A_{\alpha}(Q^{\ast})K^B_{\beta}(Q^{\ast})
  {\tilde \Gamma}^C_{AB}(Q^{\ast})
    \]
  with
\[
{\tilde \Gamma}^C_{AB}(Q^{\ast})=
\frac12\
 G^{CE}(Q^{\ast})\Bigl(\frac{\partial}
 {\partial {Q^{\ast}}^A}G_{EB}(Q^{\ast})+
 \frac{\partial}
 {\partial {Q^{\ast}}^B}G_{EA}(Q^{\ast})-
 \frac{\partial}
 {\partial {Q^{\ast}}^E}G_{AB}(Q^{\ast})\Bigr).
\]
Note also that in equation (\ref{sde_a}), 
${\Lambda}^{\alpha}_B=({\Phi}^{-1})^{\alpha}_{\mu}
{\chi}^{\mu}_B$,  ${\Lambda}^{\beta}_{AM}=
\frac{\partial}{\partial Q^*{}^M}
  \bigl({\Lambda}^{\beta}_A\bigr)$, and
 $K^B_{\sigma P}=\frac{\partial}{\partial Q^*{}^P}
 (K^B_{\sigma})$.

The superposition of the local semigroup ${\tilde U}_{\zeta}$, together with a subsequent limiting procedure,  gives the global semigroup determined  on the submanifold $\Sigma$. 

Our next transformation in the path integral reduction procedure, performed in \cite{Storchak_2}, was related to the factorization of the path integral measure generated by the process $\zeta_t$.  First of all, it  was done in the path integrals for the local evolution semigroups. In each semigroup, we have separated the local evolution, given on the orbit of the group action,   from the evolution on the orbit space. 
Then, we extended the factorization  onto the global semigroup by taking an appropriate  limit in  the  superposition of new-obtained  local semigroups.

In  case of the reduction onto non-zero momentum level,  that is when $\lambda\neq0$, it have led us to the integral relation between the path integrals for the Green's functions defined on the global  manifolds $\Sigma$ and $\cal P$:
\begin{equation}
G^{\lambda}_{pq}(Q^{*}_b,t_b;
Q^{*}_a,t_a)=
\displaystyle\int _{\cal G}G_{\cal P}(p_b\theta,t_b;
p_a,t_a) 
D_{qp}^\lambda (\theta )d\mu (\theta ),\;\;\;(Q^{*}=\pi _{\Sigma}(p)).
\label{intrelation_1}
\end{equation}
Here $D^{\lambda}_{pq}(a)$ are 
    the matrix elements  of an irreducible representation
    $T^{\lambda}$ of a group $\cal G$:
 $\sum_qD_{pq}^\lambda
(a)D_{qn}^\lambda (b)=D_{pn}^\lambda (ab)$.

The Green's function ${G}_{\cal P}(Q_b,t_b;Q_a,t_a)$ is defined\footnote{We have assumed that equation (\ref{1}) has a fundamental solution.} by semigroup (\ref{2}):
\[
 \psi(Q_a,t_a)=\int {G}_{\cal P}(Q_b,t_b;Q_a,t_a)\, \varphi_0(Q_b)\,dv_{\cal P}(Q_b)
\]
($dv_{\cal P}(Q)=\sqrt{G(Q)}\, dQ^1\cdot\dots\cdot dQ^{N_{\cal P}})$.

 The probability representation of the kernel ${G}_{\cal P}(Q_b,t_b;Q_a,t_a)$ of the semigroup (\ref{2}) (the path integral for ${G}_{\cal P}$) may be   obtained from the path integral (\ref{2}) by choosing  $\varphi_0(Q)=G^{-1/2}(Q)\,\delta (Q-Q')$ as an initial function.

 The Green's function  $G^{\lambda}_{pq}$
 is presented by the following path integral 
\begin{eqnarray}
&&G^{\lambda}_{pq}(Q^{*}_b,t_b;
Q^{*}_a,t_a)=
\nonumber\\
&&{\tilde {\rm E}}_{{\xi _{\Sigma} (t_a)=
Q^{*}_a}\atop
{\xi _{\Sigma}(t_b)=Q^{*}_b}}
\left[(\overleftarrow{\exp })_{mn}^\lambda 
(\xi _{\Sigma}(t),t_b,t_a)
\exp \left\{\frac 1{\mu ^2\kappa m}
\int_{t_a}^{t_b}\tilde{V}(\xi _{\Sigma}(u))
du\right\}\right]
\nonumber\\
&&=\int\limits_{{\xi _{\Sigma}(t_a)=Q^{*}_a
\atop  
{\xi _{\Sigma}(t_b)=Q^{*}_b}}}
d{\mu}^{{\xi}_{\Sigma}}
\exp \left\{\frac 1{\mu ^2\kappa
m}\int_{t_a}^{t_b}\tilde{V}(\xi _{\Sigma}(u))
du\right\}
\nonumber\\
&&\times\overleftarrow{\exp}
\int_{t_a}^{t_b}\Bigl\{\frac 12{\mu}^2\kappa
\Bigl[{\gamma }^{\sigma
\nu }(\xi _{\Sigma}(u))(J_\sigma
)_{pr}^\lambda (J_\nu )_{rq}^\lambda 
\nonumber\\
&&-\bigl(
G^{RS}
\tilde{\Gamma}^B_{RS}
{\Lambda}^{\beta}_B+
G^{RP}{\Lambda}^{\sigma}_R{\Lambda}^{\beta}_B
K^B_{\sigma P}-G^{CA}N^M_C
{\Lambda}^{\beta}_{AM}\bigr)
\,\,(J_\beta)_{pq}^\lambda \Bigr]du
\nonumber\\
&&+\mu\sqrt{\kappa}{\Lambda}^{\beta}_C(J_\beta)_{pq}^\lambda 
{\Pi}^C_K\tilde{\mathfrak X}^K_{\bar M}dw^{\bar M}(u)
\Bigr\}.
\label{fgreen_1}
\end{eqnarray}
 The measure in this path integral is generated by the global stochastic process ${\xi}_{{\Sigma}}(t)$ given 
on the submanifold $\Sigma$. This process is described locally by  equations (\ref{sde_q}).
 
In equation (\ref{fgreen_1}),  $\overleftarrow{\exp}(...)_{pq}^\lambda$ is a multiplicative stochastic integral. 
 This integral is a limit of the sequence of time--ordered multipliers that have been obtained as a result of breaking of a time interval $[s,t]$, $[s=t_0\le t_1
\ldots \le t_n=t]$.
The time order of these
multipliers is indicated by the arrow directed 
to the multipliers given at greater times.
We note that, by definition, a multiplicative stochastic integral represents the solution of the linear matrix stochastic differential equation.

On the right-hand side of (\ref{fgreen_1}), by  $\left.(J_\mu )_{pq}^\lambda \equiv (\frac{\partial
D_{pq}^\lambda (a)}
{\partial a^\mu })\right|_{a=e}$ we denoted  the infinitesimal
generators of the representation $D^{\lambda}(a)$:
\[
\bar{L}_\mu D_{pq}^\lambda (a)=\sum_{q^{\prime
}}(J_\mu )_{pq^{\prime
}}^\lambda D_{q^{\prime }q}^\lambda (a)
\]
($\bar{L}_\mu={\bar v}^{\alpha}_{\beta}(a)\frac{\partial}{\partial a^{\mu}}$ is a right-invariant vector field).

 The differential generator (the Hamiltonian operator) of  the matrix semigroup with the kernel (\ref{fgreen_1}) is 
 \begin{eqnarray}
 &&\frac12\mu ^2\kappa \left\{\left[
 G^{CD}N^A_CN^B_D\frac{{\partial}^2}{\partial
 Q^{*}{}^A\partial
 Q^{*}{}^B}-G^{CD}N^E_CN^M_D\,{}^H{\Gamma}^A_
 {EM}\frac{\partial}{\partial
 Q^*{}^A}
 \right.\right.
 \nonumber\\
 &&+\left.2\left(j^A_{\Roman1}+j^A_{\Roman
 2}\right)\frac{\partial}{\partial Q^*{}^A}
 +\frac{2 {\tilde V}}{(\mu ^2\kappa )^2 m}\right](I^\lambda )_{pq}
 +2N^A_CG^{CP}{\Lambda}^{\alpha}_P
 (J_\alpha )_{pq}^\lambda
 \frac{\partial}{\partial Q^{*}{}^A}
 \nonumber\\
 &&-\left(G^{RS}{\tilde
 {\Gamma}}^B_{RS}{\Lambda}^{\alpha}_B+G^{RP}{\Lambda
 }^{\sigma}_R{\Lambda}^{\alpha}_BK^{B}_{\sigma P}-G^{
 CA}N^M_C
{\Lambda}^{\alpha}_{AM})\right)
 (J_\alpha )_{pq}^\lambda 
 \nonumber\\
 &&+\biggl. G^{SB}{\Lambda}^{\alpha}_B
 {\Lambda}^{\sigma}_S
 (J_\alpha)_{pq^{\prime }}^\lambda 
 (J_\sigma)_{q^{\prime }q}^\lambda \biggr\},
 \label{op2}
 \end{eqnarray} 
where $(I^\lambda )_{pq}$ is a unity matrix.
 
 The operator acts in the space of the sections
 ${\Gamma}(\Sigma,V^*_{\lambda})$
 of the associated covector bundle with the scalar product\footnote{Another form of this scalar product is as follows
 \[
(\psi _n,\psi _m)=\!\!
\int \langle \psi _n,\psi _m
{\rangle}_{V^{\ast}_{\lambda}}
\det{\Phi}^{\alpha}_{\beta}
\prod_{\alpha =1}^{N_{\cal G}}
\delta({\chi}^{\alpha}(Q^{*})){\det}^{1/2}G_{AB}
 \,dQ^{*}{}^1\wedge\dots\wedge 
dQ^{*}{}^{N_{\cal P}}.
 \]}
\begin{eqnarray}
(\psi _n,\psi _m)&=&\int_{\Sigma}\langle \psi _n,\psi _m{\rangle}_
{V^{\ast}_{\lambda}}\,
{\det}^{1/2}\bigl((P_{\perp})^D_A \;G^{\rm H}_{DC}\,(P_{\perp})^C_B\bigr)\,        {\det}^{1/2} {\gamma}_{\alpha \beta}   
\nonumber\\
&&\times\, dQ^{*1}\wedge\ldots\wedge dQ^{*N_{\cal P}}.
\label{33}
\end{eqnarray}

${\Gamma}(\Sigma,V^*_{\lambda})$ is isomorphic to the space of the equivariant functions on $\cal P$. 
The isomorphism between the functions ${\tilde \psi}_n (p)$, such that 
\[
{\tilde \psi}_n (pg) =
D_{mn}^\lambda (g)
{\tilde \psi}_m (p),
\]
is given by the following equality:
$\;\;{\tilde \psi}_n (F(Q^{*},e))={\psi}_n (Q^*)$.

\section{Girsanov transformation}
In the case of the reduction onto the zero momentum level,
our  goal is to obtain the description of true evolution on the orbit space $\cal M$ in terms of the evolution given on an additional gauge surface $\Sigma$. By true evolution we mean such a diffusion on $\cal M$ which has the Laplace---Beltrami operator as a  differential generator. 

A required  correspondence between the diffusion on $\cal M$ and the diffusion on $\Sigma$ can be achieved only in that case when the stochastic process $\tilde{\xi} _{\Sigma}$ related to the diffusion on $\Sigma$ is described by the stochastic differential equations, which look as equations 
(\ref{sde_q}), but without the ``$j_{\Roman 2}$-term'' in the drift: 
\begin{equation}
dQ^{*}_t{}^{\small A}={\mu}^2\kappa\biggl(-\frac12
G^{EM}N^C_EN^B_M\,{}^H{\Gamma}^A_{CB}+
j^{\small A}_{\Roman 1}\biggr)dt 
+\mu\sqrt{\kappa}
N^A_C\tilde{\mathfrak X}^C_{\bar M}dw_t^{\bar M}.
\label{82}
\end{equation}

Note that in case of the reduction onto the  zero-momentum level, the differential generator of the process $\tilde{\xi} _{\Sigma}$ could be transformed into the Laplace---Beltrami operator (a differential generator of the process on $\cal M$), if we  succeded in finding the independent variables that parametrize $\Sigma$.

In the same way, in order to come to the correct description of the reduced diffusion on $\cal M$ for the reduction  onto the non-zero momentum level, we should properly transform  the  semigroup, given by the kernel (\ref{fgreen_1}). 

In the path integral (\ref{fgreen_1}), such a transformation,  in which we perform the transition 
to   the process $\tilde{\xi} _{\Sigma}$ with the local stochastic differential equations (\ref{82}) 
from the process
$\xi _{\Sigma}$ defined by the equation (\ref{sde_q}),
is known as  
the Girsanov transformation.
In spite of the fact that in the equations  (\ref{fgreen_1}) and (\ref{82}), the diffusion coefficients 
are degenerated,  the Girsanov transformation formula can be nevertheless derived  by making use of the It\^o's differentiation formula for the
composite  function. It is  necessary only to take into account the predefined
  ambiguities, which exist in the problem.


When we deal with the system of the linear parabolic differential equations, as in our case, the multiplicative stochastic integral should be also involved in the Girsanov transformation. Assuming a new form of this integral for the process $\tilde{\xi} _{\Sigma}$, we compare the differential generators for the processes  
$\xi _{\Sigma}$ and $\tilde{\xi} _{\Sigma}$. The existence and uniqueness solution
theorem for the the system of the differential equations allows us to determine a new multiplicative stochastic integral for the process $\tilde{\xi} _{\Sigma}$.

After  lengthy calculation which we omit for brevity and because of  its resemblance to the calculation
performed in \cite{Storchak_2,Storchak_3} for $\lambda=0$ case,
we come to the following expression for the 
multiplicative stochastic integral:
\begin{eqnarray}
&&
\overleftarrow{\exp}(...)^{\lambda}_{pq}(\tilde{\xi}_{\Sigma}(t))=
\overleftarrow{\exp}
\int_{t_a}^{t}\Bigl\{\frac 12{\mu}^2\kappa
\Bigl[{\gamma }^{\sigma
\nu }\,
(J_\sigma
)_{pr}^\lambda (J_\nu )_{rq}^\lambda 
\nonumber\\
&&
\;\;\;\;\;\;\;\;\;\;\;\;\;\;\;\;\;\;
-\,
G^{\rm H}_{LK}(P_{\bot})^L_A(P_{\bot})^K_E
j^A_{\Roman 2}
j^E_{\Roman 2}{I}^{\lambda}_{pq}
-2{\Pi}^C_Lj^L_{\Roman 2}{\Lambda}^{\alpha}_C(J_\alpha )_{pq}^\lambda
\nonumber\\
&&\;\;\;\;\;-\Bigl(
G^{RS}
\tilde{\Gamma}^B_{RS}
{\Lambda}^{\beta}_B+
G^{RP}{\Lambda}^{\sigma}_R{\Lambda}^{\beta}_B
K^B_{\sigma P}-G^{CA}N^M_C
{\Lambda}^{\beta}_{A,M}\Bigr)
(J_\beta)_{pq}^\lambda \Bigr]du
\nonumber\\
&&\;\;\;\;\;\;\;\;
+\mu\sqrt{\kappa}\Bigl[G^{\rm H}_{KL}\,({P}_{\bot})^L_A\,j^A_{\Roman 2}\,{ I}^{\lambda}_{pq}+{\Pi}^C_K\,
{\Lambda}^{\beta}_C\,(J_\beta)_{pq}^\lambda \,
\Bigr]\tilde{\mathfrak X}^K_{\bar M}dw^{\bar M}_u
\Bigr\}.
\label{girs}
\end{eqnarray}

We note that terms that are  proportional to ${I}^{\lambda}_{pq}$ can be factor out of the multiplicative stochastic integral.
Hence the right-hand side of (\ref{girs}) can be presented as a  product of two factors:

\begin{eqnarray}
&&
\overleftarrow{\exp}(...)^{\lambda}_{pq}(\tilde{\xi}_{\Sigma}(t))=
\exp\int^t_{t_a}\left[
-\frac12{\mu}^2\kappa \left((P_{\bot})^L_A
G^H_{LK}(P_{\bot})^K_E\right)
j^A_{\Roman 2}
j^E_{\Roman 2}du
\right.
\nonumber\\
&&\left.
\;\;\;\;\;\;\;\;\;\;\;\;\;\;\;\;\;\;\;\;\;+\mu\sqrt{\kappa}G^H_{LK}(P_{\bot})^L_A
j^A_{\Roman 2}\tilde{\mathfrak X}^K_{\bar M}dw^{\bar M}_u
\right]{ I}^{\lambda}_{pq'}
\nonumber\\
&&\times 
\overleftarrow{\exp}
\int_{t_a}^{t}\Bigl\{\frac 12{\mu}^2\kappa
\Bigl[{\gamma }^{\sigma
\nu }\,
(J_\sigma
)_{q'r}^\lambda (J_\nu )_{rq}^\lambda 
-2{\Pi}^C_Lj^L_{\Roman 2}{\Lambda}^{\alpha}_C(J_\alpha )_{q'q}^\lambda
\nonumber\\
&&\;\;\;\;\;-\Bigl(
G^{RS}
\tilde{\Gamma}^B_{RS}
{\Lambda}^{\beta}_B+
G^{RP}{\Lambda}^{\sigma}_R{\Lambda}^{\beta}_B
K^B_{\sigma P}-G^{CA}N^M_C
{\Lambda}^{\beta}_{A,M}\Bigr)
(J_\beta)_{q'q}^\lambda \Bigr]du
\nonumber\\
&&\;\;\;\;\;\;\;\;
+\mu\sqrt{\kappa}\,{\Pi}^C_K
{\Lambda}^{\beta}_C\,(J_\beta)_{q'q}^\lambda \,
\tilde{\mathfrak X}^K_{\bar M}dw^{\bar M}_u
\Bigr\}.
\label{girs_fact}
\end{eqnarray}
The first factor of (\ref{girs_fact}) coinsides with the path integral reduction Jacobian for the $\lambda=0$  case.
It was obtained in \cite{Storchak_3} in the following way. 

We first rewrote the exponential of the Jacobian for getting rid of the stochastic integral: the stochastic integral was  replaced by an ordinary integral taken with respect to the time variable.
It was made with the help of  the It\^o's identity. Then it was obtained  the geometrical representation of the Jacobian: 
\begin{eqnarray}
\Bigl(\frac{{\gamma}(Q^{\ast}(t_b))}{{\gamma}(Q^{\ast}(t_a))}\Bigr)^{\frac14}
 {\exp}\Bigl\{-\frac18{\mu}^2{\kappa}\int\limits_{t_a}^{t_b}{\tilde J}dt\Bigr\}, 
\label{redjacob_1}
\end{eqnarray}
where the integrand $\tilde J$ is equal to 
\begin{equation}
 {\tilde J}=R_{\mathcal P}-{}^{\rm H}R- R_{\mathcal G}- 
\frac14{\mathcal F}^2-||j||^2.
\label{tilde_J}
\end{equation}

In this expression, $R_{\mathcal P}$ is a scalar curvature of the original manifold $\mathcal P$.  ${}^{\rm H}R$ is a scalar curvature of the manifold  with the degenerated metric
$G^{\rm H}_{AB}$. More exactly,
\[{}^{\rm H}R\equiv G^{A' C'}N^S_{A'}N^C_{C'}N^E_M\,{}^{\rm H}R_{SEC}^{\;\;\;\;\;\;\;\;M},\]
where
$
  N^S_AN^E_M\,{}^{\rm H}R^{\;\;\;\;\;\;\;M}_{SEC}
$ 
 is equal to
\[
 N^S_AN^E_M\left(\frac{\partial}{\partial Q^{\ast}{}^S}{}^{\rm H}{\Gamma}^M_{CE}-\frac{\partial}{\partial Q^{\ast}{}^E}{}^{\rm H}{\Gamma}^M_{CS}+{}^{\rm H}{\Gamma}^K_{CE}\,{}^{\rm H}{\Gamma}^M_{KS}-{}^{\rm H}{\Gamma}^P_{CS}\,{}^{\rm H}{\Gamma}^M_{PE}\right).
\]
$R_{\mathrm {\mathcal G}}$ is the scalar curvature of the orbit:
\[
R_{\mathrm {\mathcal G}}\equiv\frac12{\gamma}^{\mu\nu} c^{\sigma}_{\mu \alpha} c^{\alpha}_{\nu\sigma}+
\frac14 {\gamma}_{\mu\sigma}{\gamma}^{\alpha\beta}{\gamma}^{\epsilon\nu}
c^{\mu}_{\epsilon \alpha}c^{\sigma}_{\nu \beta}.\]
By ${\mathcal F}^2$ we denote the following expression:
\[{\mathcal F}^2\equiv
\bigl(G^{ES}N^F_SN^B_E\bigr)\,\bigl(G^{MQ}N^P_MN^A_Q\bigr)\,{ \gamma}_{\mu \nu}\,{ \mathcal F}^{\mu}_{PF}{ \mathcal F}^{\nu}_{AB},\]
in which 
the curvature ${\mathcal F}^{\alpha}_{EP}$ of the connection  
 ${\mathscr A}^{\nu}_P={\gamma}^{\nu\mu}K^R_{\mu}\,G_{RP}$ \footnote{In the case of the reduction, this connection is naturally   defined on the principal fiber bundle.}
is given by
\[
{\mathcal F}^{\alpha}_{EP}=\displaystyle\frac{\partial}{\partial Q^{\ast}{}^E}\,{\mathscr A}^{\alpha}_P- 
\frac{\partial}{\partial {Q^{\ast}}^P}\,{\mathscr A}^{\alpha}_E
+c^{\alpha}_{\nu\sigma}\, {\mathscr A}^{\nu}_E\,
{\mathscr A}^{\sigma}_P.
\]

The last term of (\ref{tilde_J})
, the  ``square'' of the fundamental form of the orbit, 
is
\[ 
||j||^2=G^{\rm H}_{AB} \,{\gamma}^{\alpha
\mu}\,{\gamma}^{\beta\nu}\,j^A_{\alpha\beta}\,j^B_{\mu\nu}\,,
\]
where
\[
j^B_{\alpha\beta}(Q^{\ast})=
-\frac12G^{PS}N^B_PN^E_S
\,\bigl({\mathcal D}_{E}{\gamma}_{\alpha \beta}\bigr)(Q^{\ast})
\]
with 
\[
{ \mathscr D}_E{ \gamma}_{\alpha\beta}=
\Bigl(\frac{\partial}{\partial Q^{\ast}{}^E}{ \gamma}_{\alpha\beta}-c^{\sigma}_{\mu\alpha}{\mathscr A}^{\mu}_E{ \gamma}_{\sigma\beta}-c^{\sigma}_{\mu\beta}{\mathscr A}^{\mu}_E{ \gamma}_{\sigma\alpha}\,\Bigr).
\]
To obtain $j^B_{\alpha\beta}(Q^{\ast})$ we have projected 
the second fundamental form $j^C_{\alpha \beta}(Q)$ of the orbit onto the direction which is parallel with the orbit space. 
In other words, we  calculated the following expression:
${\tilde G}^{AB}{\tilde G}\left({\Pi}^C_D(Q)\bigl({\nabla}_{K_{\alpha}}K_{\beta}\bigr)^D\frac{\partial}{\partial Q^C},\frac{\partial}{\partial Q^{\ast}{}^A}\right)$,
where  $\tilde G$ was  the metric  of the manifold $\cal P$.

Therefore, the Girsanov transformation allows us to rewrite   
the integral relation (\ref{intrelation_1}) as follows
\[
\bigl({\gamma}(Q^{*}_b)\,
{\gamma}(Q^{*}_a)\bigr)^{-1/4}\,
{\tilde G}^{\lambda}_{pq}(Q^{*}_b,t_b;
Q^{*}_a,t_a)=
\displaystyle\int _{\cal G}G_{\cal P}(p_b\theta,t_b;
p_a,t_a) 
D_{qp}^\lambda (\theta )d\mu (\theta ),
\]
where the  Green's function ${\tilde G}^{\lambda}_{pq}$ 
is given by the following path integral
\begin{eqnarray}
&&{\tilde G}^{\lambda}_{pq}(Q^{*}_b,t_b;
Q^{*}_a,t_a)=
\int\limits_{{{\tilde {\tilde \xi}} _{\Sigma}(t_a)=Q^{*}_a
\atop  
{{\tilde \xi} _{\Sigma}(t_b)=Q^{*}_b}}}
d{\mu}^{{{\tilde \xi}}_{\Sigma}}
\exp \biggl\{
\int_{t_a}^{t_b}\Bigl(\frac{\tilde{V}({\tilde \xi} _{\Sigma}(u))}{\mu ^2\kappa
m}-\frac18\mu ^2\kappa
{\tilde J}\Bigr)
du\biggr\}
\nonumber\\
&&\times 
\overleftarrow{\exp}
\int_{t_a}^{t_b}\Bigl\{\frac 12{\mu}^2\kappa
\Bigl[{\gamma }^{\sigma
\nu }\,
(J_\sigma
)_{pr}^\lambda (J_\nu )_{rq}^\lambda 
-2{\Pi}^C_Lj^L_{\Roman 2}{\Lambda}^{\alpha}_C(J_\alpha )_{pq}^\lambda
\nonumber\\
&&\;\;\;\;\;-\Bigl(
G^{RS}
\tilde{\Gamma}^B_{RS}
{\Lambda}^{\beta}_B+
G^{RP}{\Lambda}^{\sigma}_R{\Lambda}^{\beta}_B
K^B_{\sigma P}-G^{CA}N^M_C
{\Lambda}^{\beta}_{A,M}\Bigr)
(J_\beta)_{pq}^\lambda \Bigr]du
\nonumber\\
&&\;\;\;\;\;\;\;\;
+\mu\sqrt{\kappa}\,{\Pi}^C_K
{\Lambda}^{\beta}_C\,(J_\beta)_{pq}^\lambda \,
\tilde{\mathfrak X}^K_{\bar M}dw^{\bar M}_u
\Bigr\}.
\label{fgreen_2}
\end{eqnarray}
 ${\tilde G}^{\lambda}_{pq}$ is the kernel of the evolution semigroup   which describes the true reduced evolution on the orbit space $\cal M$. This semigroup acts in the space of sections 
${\Gamma}(\Sigma,V^*_{\lambda})$
 of the associated covector bundle $P\times_{\mathcal G}V^*_{\lambda}$ with the following scalar product:
\begin{eqnarray}
(\psi _n,\psi _m)&=&\int_{\Sigma}\langle \psi _n,\psi _m{\rangle}_
{V^{\ast}_{\lambda}}\,
{\det}^{1/2}\bigl((P_{\perp})^D_A \;G^{\rm H}_{DC}\,(P_{\perp})^C_B\bigr)\,    
\nonumber\\
&&\times\, dQ^{*1}\wedge\ldots\wedge dQ^{*N_{\cal P}}.
\label{scal_product}
\end{eqnarray}

The differential generator of  the matrix semigroup with the kernel ${\tilde G}^{\lambda}_{pq}$ is 
 \begin{eqnarray}
 &&\frac12\mu ^2\kappa \left\{\left[
 G^{CD}N^A_CN^B_D\frac{{\partial}^2}{\partial
 Q^{*}{}^A\partial
 Q^{*}{}^B}-G^{CD}N^E_CN^M_D\,{}^H{\Gamma}^A_
 {EM}\frac{\partial}{\partial
 Q^*{}^A}
 \right.\right.
 \nonumber\\
 &&+\left.2j^A_{\Roman1}\frac{\partial}{\partial Q^*{}^A}
 +\frac{2 {\tilde V}}{(\mu ^2\kappa )^2 m}-\frac14{\tilde J} \right](I^\lambda )_{pq}
 +2N^A_CG^{CP}{\Lambda}^{\alpha}_P
 (J_\alpha )_{pq}^\lambda
 \frac{\partial}{\partial Q^{*}{}^A}
 \nonumber\\
 &&-\left(G^{RS}{\tilde
 {\Gamma}}^B_{RS}{\Lambda}^{\alpha}_B+G^{RP}{\Lambda
 }^{\sigma}_R{\Lambda}^{\alpha}_BK^{B}_{\sigma P}-G^{
 CA}N^M_C
{\Lambda}^{\alpha}_{AM}\right)(J_\alpha )_{pq}^\lambda 
 \nonumber\\
 &&+{\Lambda}^{\alpha}_C
 {\gamma}^{\mu\nu}[{\triangledown}_{K_{\mu}}K_{\nu}]^C(J_\alpha )_{pq}^\lambda +\biggl. G^{SB}{\Lambda}^{\alpha}_B
 {\Lambda}^{\sigma}_S
 (J_\alpha)_{pq^{\prime }}^\lambda 
 (J_\sigma)_{q^{\prime }q}^\lambda \biggr\}.
 \label{operator_2}
 \end{eqnarray} 
The first term of the last line in (\ref{operator_2})  comes  from $(-2{\Pi}^C_Lj^L_{\Roman 2}{\Lambda}^{\alpha}_C$) - term of the multiplicative stochastic integral given in (\ref{fgreen_2}). Its derivation is based on the following relations:
\[
{\Pi}^R_C {\Lambda}^{\alpha}_R={\Lambda}^{\alpha}_C-
{\mathscr A}^{\alpha}_C, \quad {\mathscr A}^{\alpha}_C\,
{\gamma}^{\mu\nu}[{\triangledown}_{K_{\mu}}K_{\nu}]^C=0.
\]

In the next section we will obtain  another representation for the multiplicative stochastic integral.
For this purpose,  it is sufficient 
to consider 
the transformation of  the differential operator  (\ref{operator_2}), since there exists  
 a quite definite relationship 
 between the 
integrand of the path integral and the corresponding differential generator.

\section{The horizontal Laplacian}
It is well-known that the horizontal Laplacian ${\triangle}^{{\mathcal E}}$
\begin{eqnarray*}
\left({\triangle}^{{\mathcal E}}\right)^{\lambda}_{pq}
&=&{\sum}_{\bar k=1}^{n_{\cal M}}
\left({\nabla}^{\mathcal E}_{X^i_{\bar k}{\rm e_i}}
{\nabla}^{\mathcal E}_{X^j_{\bar k}{\rm e_j}}-
{\nabla}^{\mathcal E}_{{\nabla}^{\mathcal M}_{X^i_{\bar k}{\rm e_i}} {X^j_{\bar k}{\rm e_j}}}\right)^{\!\!\lambda}_{\!\!pq}\nonumber\\
&=&{\triangle}_{\cal M}\,
{\rm I}^{\lambda}_{pq} 
+2h^{ij}({\rm {\Gamma}^{\mathcal E}})^{\lambda}_{ipq}\,{\partial}_j\nonumber\\
&&-\,h^{ij}\left[{\partial}_i ({\rm {\Gamma}^{\mathcal E}})^{\lambda}_{jpq}
-({\rm {\Gamma}^{\mathcal E}})^{\lambda}_{ip{q}^{'}}({\rm {\Gamma}^{\mathcal E}})^{\lambda}_{j{q}^{'}q}+({\rm {\Gamma}^{\cal M}})^{m}_{ij}\,
({\rm{\Gamma}}^{\mathcal E})^{\lambda}_{mpq}\right],
\end{eqnarray*}
determined on the space of the sections of the associated vector bundle $\mathcal E=P\times_{\mathcal G}V_{\lambda}$, 
is an invariant operator which  can be considered as  a generalization of  
the Lapalace--Beltrami operator given on  the base manifold  $\mathcal M$. 
It would be naturally to expect that 
in the case of description of the evolution by means of  dependent variables,
there is also a  corresponding operators which may be refer to as the horizontal Laplacian.

For the covector bundle, such an operator may be given by the following expression:
\begin{eqnarray*}
&&\!\!\!\!\!\left({\triangle}^{{\mathcal E^*}}\right)^{\lambda}_{pq}
={\sum}_{\bar{\scriptscriptstyle M}=1}^{n_{\cal P}}
\left({\nabla}^{\mathcal E^*}_{Y^A_{\bar M}{\rm e_A}}
{\nabla}^{\mathcal E^*}_{Y^B_{\bar M}{\rm e_B}}-
{\nabla}^{\mathcal E^*}_{{\nabla}^{\mathcal M}_{Y^A_{\bar M}{\rm e_A}} {Y^B_{\bar M}{\rm e_B}}}\right)^{\!\!\lambda}_{\!\!pq}\nonumber\\
&&\!\!\!\!\!\!={\triangle}_{\cal M}\,
{\rm I}^{\lambda}_{pq} 
-2\,G^{LM}N^E_LN^C_M\,({\rm {\Gamma}^{\mathcal E}})^{\lambda}_{Epq}\,{\partial}_{Q^{*C}}\nonumber\\
&&\!\!\!\!\!\!\!\!-G^{LM}N^E_LN^B_M\left[{\partial}_{Q^*{}^E} (N^C_B({\rm {\Gamma}^{\mathcal E}})^{\lambda}_{Cpq})
-({\rm {\Gamma}^{\mathcal E}})^{\lambda}_{E\,p{q}^{'}}({\rm {\Gamma}^{\mathcal E}})^{\lambda}_{B{q}^{'}q}-{}^{\rm H}{\rm {\Gamma}}^{C}_{EB}N^D_C
({\rm{\Gamma}}^{\mathcal E})^{\lambda}_{Dpq}\right],
\end{eqnarray*}
in which ${Y}^A_{\bar M}=N^A_P{\mathfrak X}^P_{\bar M}$ 
is defined by the  equality  
$\sum^{n_{P}}_{\bar{
{\scriptscriptstyle M}}
\scriptscriptstyle =1}
Y_{\bar{M}}^AY_
{\bar{M}}^B=G^{PR}N^A_{P}N^B_{R}$ and where 
$({\rm{\Gamma}}^{\mathcal E})^{\lambda}_{Bpq}={\mathscr A}^{\alpha}_B\,(J_{\alpha})^{\lambda}_{pq}$.

The covariant derivative  ${\nabla}^{{\pi}^*}$ is defined as 
\[
 {\nabla}^{{\pi}^*}_{{\rm e}_B}u_p=N^D_B\left({\rm I}^{\lambda}_{pq}\frac{\partial }{\partial Q^{*D}}-{\mathscr A}^{\alpha}_D(J_{\alpha})^{\lambda}_{pq}\right)u_q,
\]
and 
\[
{\nabla}^{\scriptscriptstyle{\mathcal M}}_{\rm e_A}\, {\rm e_B}={}^H{\Gamma}^C_{AB}\,{\rm e_C}.
\]

The 
 horizontal Laplacian ${\triangle}^{{\mathcal E^*}}$
 can be also written as follows
\begin{eqnarray}
&&G^{LM}N^E_LN^C_M\left\{\left(\frac{{\partial}^2}{{\partial Q^{*E}}{\partial Q^{*C}}}+
\frac{\partial}{\partial Q^{*C}}(N^B_E) 
\frac{\partial}{\partial Q^{*B}}-\,{}^{\scriptstyle {\rm H}}{\Gamma}^B_{EC}N^D_B\frac{\partial}{\partial Q^{*D}}\right){\rm I}^{\lambda}_{pq}\right.\nonumber\\
&&\left(-2{\mathscr A}^{\alpha}_E
\frac{\partial}{\partial Q^{*C}}-{\mathscr A}^{\alpha}_B\frac{\partial}{\partial Q^{*E}}(N^B_C)-  \frac{\partial}{\partial Q^{*E}}({\mathscr A}^{\alpha}_C)
+\,{}^{\scriptstyle {\rm H}}{\Gamma}^B_{EC}N^D_B {\mathscr A}^{\alpha}_D\right)(J_{\alpha})^{\lambda}_{pq}
\nonumber\\
&&\left.
+({\mathscr A}^{\beta}_E 
J_{\beta})^{\lambda}_{pq^{'}}
({\mathscr A}^{\alpha}_CJ_{\alpha})^{\lambda}_{q^{'}q}\right\}.
\label{operator_3}
\end{eqnarray}
It turns out, that operator (\ref{operator_3}) is intrinsically
related to the the operator (\ref{operator_2}).

First note that  diagonal parts of these operator  
(without taking into account the potential terms $\tilde V$ and $\tilde J$ in (\ref{operator_2})) 
are equal. 
It may be checked with the help of the following identity
\begin{eqnarray*}
&&-\frac12N^A_{A'}\,{}^{\scriptstyle {\rm H}}{\Gamma}^{A'}_{CD}\,N^C_{C'}N^D_{D'}G^{C'D'}+\frac12
N^A_{LM}\,N^L_{L'}N^M_{M'}\,G^{L'M'}
\nonumber\\
&&=-\frac12G^{EM}N^C_EN^B_M\,{}^{\scriptstyle {\rm H}}{\Gamma}^{A}_{CB}+j^A_{\Roman 1}.
\nonumber
\end{eqnarray*}

The off-diagonal 
matrix elements
of the operators (\ref{operator_2}) and (\ref{operator_3}), 
that include the  generator $(J_{\alpha})^{\lambda}_{pq}$,  are also equal. 
In order to show this, in the     operator (\ref{operator_2}), one should rewrite such  terms   in the following way
\begin{equation} 
-\left(^{\perp}G^{RS}{\tilde\Gamma}^P_{RS}{\Lambda}^{\alpha}_P
+G^{RP}{\Lambda}^{\sigma}_R{\Lambda}^{\alpha}_BK^B_{\sigma P}-G^{CA}N^M_C{\Lambda}^{\alpha}_{AM}
-{\gamma}^{\mu \sigma}{\Lambda}^{\alpha}_PK^A_{\mu}K^P_{\sigma A}\right),
\label{1_j}
\end{equation}
where $ ^{\perp}G^{RS}=G^{RS}-K^R_{\mu}{\gamma}^{\mu \nu}K^S_
{\nu}$,
and the analagous terms of the operator (\ref{operator_3}) as follows 
\begin{equation} 
-G^{PQ}N^E_PN^B_QN^C_{BE}
{\mathscr  A}^{\alpha}_C-G^{PQ}N^E_PN^C_Q\frac{\partial}{\partial Q^{*E}}({\mathscr A}^{\alpha}_C)+G^{PQ}N^A_PN^C_Q \, ^{\rm H}{\Gamma}^B_{AC}N^D_B{\mathscr A}^{\alpha}_D.
\label{2_j}
\end{equation}
Replacing  the term, which involve the derivative of ${\mathscr A}^{\alpha}_C$,   with  the expression 
\begin{eqnarray*}
&&G^{LM}N^E_LN^B_M\frac{\partial}{\partial Q^{*}{}^E}\,({\mathscr A}^{\alpha}_B)=
N^E_R\,{\Gamma}^R_{ES}{\gamma}^{\alpha\sigma}K^S_{\sigma}+G^{PB}N^E_P{\gamma}^{\alpha\sigma}K^S_{\sigma}\,{\Gamma}^R_{EB}G_{RS}
\nonumber\\
&&\qquad\qquad
+N^E_P{\gamma}^{\alpha\sigma}K^P_{\sigma E}+G^{LM}N^E_L{\Lambda}^{\sigma}_M {\gamma}^{\alpha\mu}K^C_{\mu}G_{CD}K^D_{\sigma E}
\end{eqnarray*}
and making use of the identity
\begin{eqnarray*}
&&N^A_{\tilde A}\,\,{}^H{\Gamma}^{\tilde A}_{CD}\,N^C_{\tilde C}N^D_{\tilde D}\,G^{\tilde C\tilde D}=
\nonumber\\
&&\;\;N^A_{LM}N^L_{\tilde L}N^M_{\tilde M}
G^{\tilde L\tilde M}
-G^{CT}N^U_CN^A_{TU}+{}^{\bot}G^{CR}{\Lambda}^{\beta}_CN^A_TK^T_{\beta R}+{}^{\bot}G^{LM}{\tilde\Gamma}^D_{LM}N^A_D,
\end{eqnarray*}
one can arrive at the equality of the  transformed expressions. It will be noted that in the expression obtained as a result of the transformation of
(\ref{2_j}), besides of the necessary terms, that are equal to the corresponding terms coming from (\ref{1_j}), there are redundent terms. But, it can be verified that these terms are mutually cancelled. It follows from the calculation in which one should takes into account the Killing identity,  the equality
\[
 {\gamma}^{\beta \nu}({\triangle}_{K_{\nu}}K_{\beta})^P{\mathscr A}^{\alpha}_P=0,
\]
which is obtained  from the identity
\[
-{\gamma}^{\beta \nu}({\triangle}_{K_{\nu}}K_{\beta})^T=\frac12G^{PT}N^E_P\;\Bigl({\gamma}^{\mu \nu}\frac{\partial}{\partial Q^{* E}}{\gamma}_{\mu \nu}\Bigr),
\]
and the condition $c^{\alpha}_{\beta \alpha}=0$, which is valid for the structure constants of the semisimple Lie groups.

Except for the potential terms, the only distinction 
between (\ref{operator_2}) and (\ref{operator_3}) consists of the terms that involve the product of two  group generators.  But, since 
\[ 
G^{LM}N^E_LN^P_M\,{\mathscr A}^{\mu}_E{\mathscr A}^{\nu}_P=G^{EP}{\Lambda}^{\mu}_E{\Lambda}^{\nu}_P-{\gamma}^{\mu\nu},
\]
we can present the operator (\ref{operator_2}) as
\[
\frac12{\mu}^2{\kappa}\left[\bigl({\triangle}^{{\mathcal E^*}}\bigr)_{pq}^{\lambda}+{\gamma}^{\mu \nu}
(J_\mu)_{pq^{\prime }}^\lambda 
 (J_\nu)_{q^{\prime }q}^\lambda \right]+\left(
\frac{1}{\mu ^2\kappa  m} {\tilde V}-\frac18\mu ^2\kappa  {\tilde J} \right)(I^\lambda )_{pq}\,.
\]

\section{The path integral for the matrix Green's function ${\tilde G}^{\lambda}_{pq}$}
Now we can rewrite the multiplicative stochastic integral in  the path integral (\ref{fgreen_2}). We already know that  $(J_\alpha)_{pq}^\lambda$--terms of the drift in the integrand of the multiplicative stochastic integral are equal to the corresponding terms (\ref{2_j}) of the operator (\ref{operator_3}).
These terms can be rewritten  as follows
\begin{eqnarray*}
&&-G^{PQ}N^A_PN^B_QN^E_AN^C_{B,E}
{\mathscr  A}^{\alpha}_C-G^{PQ}N^A_PN^B_QN^C_BN^E_A
\frac{\partial}{\partial Q^{*E}}
({\mathscr A}^{\alpha}_{C})
\nonumber\\
&&\;\;\;\;+G^{PQ}N^A_PN^C_Q \, ^{\rm H}{\Gamma}^B_{AC}N^D_B{\mathscr A}^{\alpha}_D,
 \end{eqnarray*}
and also as
\[
 -G^{PQ}N^E_PN^B_Q\,{\nabla}^{\rm H}_{{\rm e}_E}(N^C_B{\mathscr A}^{\alpha}_C).
\]

The coefficient ${\Pi}^C_K{\Lambda}^{\beta}_C$ of the diffusion term of the integrand may be written in the form 
\[
 {\Pi}^C_K{\Lambda}^{\beta}_C={\Lambda}^{\beta}_K-{\mathscr A}^{\beta}_K.
\]
Thus, we obtain the following path integral representation of the matrix Green's function ${\tilde G}^{\lambda}_{pq}$:
\begin{eqnarray}
&&{\tilde G}^{\lambda}_{pq}(Q^{*}_b,t_b;
Q^{*}_a,t_a)=
\int\limits_{{{\tilde {\tilde \xi}} _{\Sigma}(t_a)=Q^{*}_a
\atop  
{{\tilde \xi} _{\Sigma}(t_b)=Q^{*}_b}}}
d{\mu}^{{{\tilde \xi}}_{\Sigma}}
\exp \biggl\{
\int_{t_a}^{t_b}\Bigl(\frac{\tilde{V}
}{\mu ^2\kappa
m}-\frac18\mu ^2\kappa
{\tilde J}\Bigr)
du\biggr\}
\nonumber\\
&&\times 
\overleftarrow{\exp}
\int_{t_a}^{t_b}\Bigl\{\frac 12{\mu}^2\kappa
\Bigl[{\gamma }^{\sigma
\nu }\,
(J_\sigma
)_{pr}^\lambda (J_\nu )_{rq}^\lambda 
-G^{PQ}N^E_PN^B_Q\,{\nabla}^{\rm H}_{{\rm e}_E}(N^C_B{\mathscr A}^{\alpha}_C)
(J_\alpha )_{pq}^\lambda  \Bigr]du
\nonumber\\
&&\;\;\;\;\;\;\;\;
-\mu\sqrt{\kappa}\,N^B_K{\mathscr A}^{\alpha}_B
\,(J_\alpha)_{pq}^\lambda \,
\tilde{\mathfrak X}^K_{\bar M}dw^{\bar M}_u
\Bigr\}.
\label{fgreen_3}
\end{eqnarray} 
In $(Q^*_b,t_b)$-variables this Green's function satisfies the forward Kolmogorov equation with the operator 
\[
{\hat H}_{\kappa}=\frac{\hbar\kappa}{2m}\left[({\triangle}^{\mathcal E})^{\lambda}_{pq}+ {\gamma}^{\mu \nu}
(J_\mu)_{pq^{\prime }}^\lambda 
 (J_\nu)_{q^{\prime }q}^\lambda \right]-\frac{\hbar \kappa}{8m}[\tilde J\,]I^\lambda _{pq}+
 \frac{\tilde V}{\hbar \kappa}I^\lambda _{pq},
\]
where  the horizontal Laplacian $({\triangle}^{{\mathcal E}})^{\lambda}_{pq}$ is
\begin{eqnarray*}
&&\!\!\!\!\!\left({\triangle}^{{\mathcal E}}\right)^{\lambda}_{pq}
={\sum}_{\bar{\scriptscriptstyle M}=1}^{n_{\cal P}}
\left({\nabla}^{\mathcal E}_{Y^A_{\bar M}{\rm e_A}}
{\nabla}^{\mathcal E}_{Y^B_{\bar M}{\rm e_B}}-
{\nabla}^{\mathcal E^*}_{{\nabla}^{\mathcal M}_{Y^A_{\bar M}{\rm e_A}} {Y^B_{\bar M}{\rm e_B}}}\right)^{\!\!\lambda}_{\!\!pq}\nonumber\\
&&\!\!\!\!\!\!={\triangle}_{\cal M}\,
{\rm I}^{\lambda}_{pq} 
+2\,G^{LM}N^E_LN^C_M\,({\rm {\Gamma}^{\mathcal E}})^{\lambda}_{Epq}\,{\partial}_{Q^{*C}}\nonumber\\
&&\!\!\!\!\!\!\!\!-G^{LM}N^E_LN^B_M\left[{\partial}_{Q^*{}^E} (N^C_B({\rm {\Gamma}^{\mathcal E}})^{\lambda}_{Cpq})
-({\rm {\Gamma}^{\mathcal E}})^{\lambda}_{E\,p{q}^{'}}({\rm {\Gamma}^{\mathcal E}})^{\lambda}_{B{q}^{'}q}+{}^{\rm H}{\rm {\Gamma}}^{C}_{EB}N^D_C
({\rm{\Gamma}}^{\mathcal E})^{\lambda}_{Dpq}\right].
\end{eqnarray*}
The Laplace operator  ${\triangle}_{\mathcal M}$ is
\[
{\triangle}_{\mathcal M}= G^{CD}N^A_CN^B_D\frac{{\partial}^2}{\partial
 Q^{*}{}^A\partial
 Q^{*}{}^B}-G^{CD}N^E_CN^M_D\,{}^H{\Gamma}^A_
 {EM}\frac{\partial}{\partial
 Q^*{}^A}
 +2j^A_{\Roman1}\frac{\partial}{\partial Q^*{}^A}.
 \]

At $\kappa=i$ the forward Kolmogorov equation becomes the Schr\"odinger equation with the Hamilton operator $\hat H_{\mathcal E}=-\frac{\hbar}{\kappa}{\hat H}_{\kappa}|_{\kappa=i}$. The operator $\hat H_{\mathcal E}$ acts in the Hilbert space of the sections of the associated vector bundle ${\mathcal E}=P\times_{\mathcal G}V_{\lambda}$. The scalar product in this space has the same volume measure as in (\ref{scal_product}).

\section{Conclusion}
In this paper, we have considered the transformation of the path integral obtained as a result of the reduction of the finite-dimensional dynamical system with a symmetry. We have dealt with the reduction, which in the constrained dynamical system theory  is called the reduction onto the non-zero momentum level.

Because of exploiting the dependent variables for the description of the local reduced motion, we were forced to consider only the trivial principal fiber bundles. Thereby, our consideration is  a global one  only for the trivial principal bundle. For the nontrivial principal fiber bundle, that may be related to the dynamical system with a symmetry, the dependent variable description of the evolution is valid in a some  local domain.  

Although for the nontrivial principal fiber bundles, there is a method  \cite{Kelnhofer} which allows us
to extend the local evolution to a global one, but in general this problem remains unsolved,  especially for the reason of a possible existence of the non-trivial topology of the orbit space.

In conclusion, we note that besides of the application of
the obtained path  integral representation (and the integral relation) in
 the quantization of the finite-dimensional dynamical systems with a symmetry, this representation  may be useful for a quantum description (in the Schr\"odinger's  approach) of the excited modes in the gauge fields models.


\begin{thebibliography}{**}
\bibitem{Tanim}
 Landsman N P and Linden N 1991
{\it Nucl. Phys.} {\bf B365} 121;\\
Tanimura S and  Tsutsui I 1995
{\it Mod. Phys. Lett.} {\bf A34} 2607;\\
McMullan D and Tsutsui I 1995 {\it Ann. Phys.} 
{\bf 237} 269. 

\bibitem{Kunsht}
Kunstatter G 1992
{\it Class. Quant.Grav.} {\bf 9} 1466-86.
 

\bibitem{Falk}
Falck N K and  Hirshfeld A C 1982
{\it Ann. Phys.} {\bf 144}  34;\\ 
Gavedzki K 1982
{\it Phys.Rev.} {\bf D26} 3593.

\bibitem{Lott}
Lott J 1984
{\it Comm. Math. Phys.} {\bf 95}  289.

\bibitem{Faddeev} 
L. D. Faddeev, {\it Teor. i Mat. Fyz.} {\bf 1} (1969) 3 (in
Russian);\\
L. D. Faddeev, V. N. Popov, {\it Phys. Lett.} 
{\bf 25B} (1967) 30.

\bibitem{Storchak_1} 
S. N. Storchak, {\it J. Phys. A: Math. Gen.}
{\bf 34} (2001) 9329,\\
IHEP Preprint 96-110, Protvino, 1996;\\
S. N. Storchak. {\it Bogolubov transformation in path integral on manifold with a group action.} IHEP Preprint 98-1, Protvino, 1998;\\
S. N. Storchak, {\it Physics of Atomic Nuclei} {\bf 64} n.12 (2001) 2199 


\bibitem{Elworthy_2}
K. D. Elworthy, Y. Le Jan, Xue-Mei Li {\it The Geometry of Filtering (Preliminary Version)} (2008), arXiv:0810.2253

\bibitem{Paycha}
M. Arnaudon, S. Paycha, {\it Stochastic and Stochastic Reports} {\bf 53} (1995) 81.




\bibitem{Storchak_2}
S. N. Storchak, {\it J. Phys. A: Math. Gen.} 
{\bf 37} (2004) 7019,\\
IHEP Preprint 2000-54, Protvino, 2000; arXiv:math-ph/0311038

\bibitem{Daletskii}
Ya. I. Belopolskaya, Yu. L. Daletskii, 
{\it Russ. Math. Surveys} {\bf 37} 109 (1982);
{\it Usp. Mat. Nauk} {\bf 37}
n.3 (1982) 95  (in Russian);\\
Yu. L. Daletskii, {\it Usp. Mat. Nauk} {\bf 38} n.3 (1983) 87 (in
Russian);\\
Ya. I. Belopolskaya and Yu. L. Daletskii,
{\it Stochastic equations and differential geometry}
(Kluwer, Dordrecht, 1990),
Mathematics and Its Applications, Soviet Series, 30. 

\bibitem{Storchak_3}
S. N. Storchak, {\it J. of Geometry and Physics} 
{\bf 59} (2009) 1155.\\


\bibitem{Kelnhofer}
H. H\"uffel, G. Kelnhofer, {\it Ann. of Phys.} 
{\bf 266} (1998) 417;\\
{\it Ann. of Phys.} {\bf 270} (1998) 231.








\end{thebibliography}
\end{document}